\documentclass[a4paper,graphics,natbib,graphicx,psfig,amssymb,ccaption]{article}
\usepackage{lscape}
\usepackage{graphicx}
\newcommand{\affil}[1]{$^{\rm #1}$}
\usepackage[authoryear]{natbib}
\bibpunct{(}{)}{;}{a}{}{,}
\date{} %Please leave the date blank

\title{\large\bf\flushleft An Improved Metallicity Calibration with UBV Photometry}
\author{\parbox{\textwidth}{\flushleft
\vspace{-0.5cm}
{\it S. Karaali\affil{A}, S. Bilir\affil{B}, S. Ak\affil{B}, E. Yaz\affil{B} and B. Co\c skuno\u glu\affil{B}}\\
\vspace{0.4cm}
{\small \affil{A}\,Beykent University, Faculty of Science and Letters, Department 
of Mathematics and Computer, Beykent Ayaza\u ga Campus, 34398, Istanbul, Turkey, Email: skaraali@beykent.edu.tr}\\
{\small \affil{B}\,Istanbul University, Faculty of Sciences, Department 
of Astronomy and Space Sciences, 34119 University, Istanbul, Turkey}\\}}
\begin{document}
\twocolumn[
\begin{changemargin}{.8cm}{.5cm}
\begin{minipage}{.9\textwidth}
\vspace{-1cm}
\maketitle
%
%
%%%%%%%%%%%%%     ABSTRACT    %%%%%%%%%%%%%
%Abstract of no more than 200 words here.
\small{{\bf Abstract:} We used the data of 701 stars covering the colour index interval $0.32<B-V\leq1.16$, with metallicities $-1.76\leq[Fe/H]\leq+0.40$ dex, which were taken from PASTEL catalogue and estimated metallicity dependent guillotine factors which provide a more accurate metallicity calibration. We reduced the metallicities of 11 authors to the metallicities of \cite{Valenti05}, thus obtained a homogeneous set of data which increased the accuracy of the calibration, i.e. $[Fe/H]=-14.316\delta^{2}_{0.6}-3.557\delta_{0.6}+0.105$. Comparison of the metallicity residuals, for two sets of data, based on the metallicity dependent guillotine factors with the ones obtained via metal free guillotine factors, shows that metallicities estimated by means of new guillotine factors are more accurate than the other ones. This advantage can be used in the metallicity gradient investigation of the Galactic components, i.e. thin disc, thick disc and halo.        
}

\medskip
\medskip
\end{minipage}
\end{changemargin}
]
\small

\section{Introduction}

\cite{Roman55} interpreted the weakness of the metallic lines in the F–- and G- type spectra by comparison of the $B–-V$ and $U–-B$ colours for each star. She stated that an ultraviolet excess, ranging from 0.0 to more than 0.2 mag, which is found in most high–-velocity stars, is well correlated with the weakness of the lines. Moreover, both anomalies are correlated with velocity, in the sense that the stars with the weakest lines also have the largest ultraviolet excess and the largest space velocities.

Following \cite{Schwarzschild55}, \cite{Sandage59} interpreted the observed ultraviolet excess for subdwarfs with the ``blanketing model''. This model predicts that the change in $B–-V$ colour index for a given observed ultraviolet excess, $\delta(U–-B)$, for F and G subdwarfs is sufficient to move most of the subdwarfs with known $M_V$ on the Hyades main sequence. The essential point of the theory is that the Fraunhofer lines affect the $U$, $B$, and $V$ regions of the spectrum in different ways so that a weakening of the lines produces changes in the observed colour indices: $U–-B$ and $B-–V$. If the relation between the effect on $U–-B$ and $B-–V$ is known, then the correction to the observed $B–-V$ can be computed from the observed  ultraviolet excess. Because the observed $B–-V$ for weak line stars will be bluer than that for strong line stars of the same temperature, the weak line stars will fall below the standard main sequence. Therefore, because of the relationship between $\Delta(U-–B)$ and  $\Delta(B-–V)$ we should expect that the displacement of a weak line star below the standard main sequence will be correlated with the observed ultraviolet excess. \cite{Wallerstein60} calibrated the ultraviolet excess in terms of $[Fe/H]$ for the first time, and \cite{Wallerstein62} improved this calibration. The scheme between the observed ultraviolet excess, $\delta(U–-B)$ and the blanketing corrections $\Delta(U-–B)$ and  $\Delta(B-–V)$ (for a hypothetical star) are given in Fig. 1. 

The shapes of the blanketing vectors in the $(U-–B)–-(B-–V)$ diagram are such that stars with different $B-–V$ values with the same metal abundance will show different ultraviolet excess values. For red stars, $\delta(U-–B)$ is partially guillotined because the blanketing line is nearly parallel to the intrinsic Hyades line. If the metal abundances are to be compared among stars of different colours, such as in the works carried out for the estimation of the metallicity gradient for the Galactic fields, corrections to the observed $\delta(U-–B)$ are needed. \cite{Wildey62} provided the basis on which normalized ultraviolet excess was computed by \cite{Sandage69} and \cite{Carney79}. \cite{Sandage69} gave a procedure to correct ultraviolet excess values for stars with the same metal abundance, but of different colours for the effect of the guillotine. He plotted 112 stars of large proper motion onto the $(U–-B)-–(B-V)$ two colour diagram and compared the $U–-B$ colours of maximum abundance with that of Hyades for the same $B-–V$ colour. The results are given in Table 1. The columns give: (1) $B–-V$ colour, (2) the Hyades fiducial line, (3) the maximum $U-–B$ value for the sample star for the same $B-–V$ of the Hyades star, (4) the $\delta(U-–B)$ ultraviolet excess of the sample star in question, and (5) the ratio of the excess at $(B–-V)=0.60$ (where $\delta(U-–B)$ is maximum), $\delta_{0.6}$, to the excess at any other $B–-V$. This ratio is defined as ``guillotine factor'' in this paper, i.e. $f_S=\delta_{0.6}/\delta(U-–B)$, where the subscript ``$S$'' refers to Sandage. Table 1 gives the guillotine factors of \cite{Sandage69} for a set of 16 colours with 0.35$\leq$$B–-V$$\leq$1.10. One can estimate guillotine factors for a larger set of $B–-V$ colours by applying an interpolation formula to the data in Table 1. This is the case in some of our works \citep{Karaali03, Ak07, Ak08, Yaz10}. 

%TABLE 1
\begin{table}
\setlength{\tabcolsep}{2pt}
\center
{\scriptsize
\caption{The guillotine factors of \cite{Sandage69}. The symbols are explained in the text.}
\begin{tabular}{ccccc}
\hline
$B-V$ & $(U-B)_{H}$ & $(U-B)_{M}$ & $\delta(U-B)$ & $\delta_{0.6}/\delta(U-B)$ \\
\hline
0.35 & 0.03 & -0.22 & 0.25 & 1.24 \\
0.40 & 0.01 & -0.25 & 0.26 & 1.19 \\
0.45 & 0.00 & -0.27 & 0.27 & 1.15 \\
0.50 & 0.03 & -0.25 & 0.28 & 1.11 \\
0.55 & 0.08 & -0.22 & 0.30 & 1.03 \\
0.60 & 0.13 & -0.18 & 0.31 & 1.00 \\
0.65 & 0.19 & -0.11 & 0.30 & 1.03 \\
0.70 & 0.25 & -0.03 & 0.28 & 1.10 \\
0.75 & 0.34 &  0.08 & 0.26 & 1.19 \\
0.80 & 0.43 &  0.19 & 0.24 & 1.29 \\
0.85 & 0.54 &  0.32 & 0.22 & 1.41 \\
0.90 & 0.64 &  0.44 & 0.20 & 1.55 \\
0.95 & 0.74 &  0.55 & 0.19 & 1.63 \\
1.00 & 0.84 &  0.67 & 0.17 & 1.82 \\
1.05 & 0.94 &  0.79 & 0.15 & 2.06 \\
1.10 & 0.99 &  0.87 & 0.12 & 2.58 \\
\hline
\end{tabular}
}  
\end{table}

%FIGURE 1
\begin{figure}
\begin{center}
\includegraphics[scale=1, angle=0]{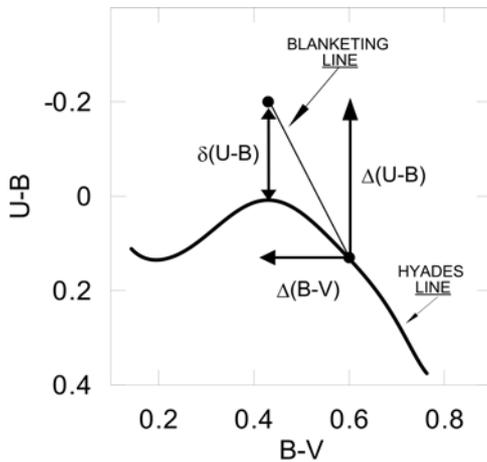}
\caption[] {The scheme between the observed ultraviolet excess, $\delta(U-B)$ and the blanketing corrections $\Delta (U-B)$ and $\Delta (B-V)$ taken from \cite{Sandage59}.}
\end{center}
\end{figure}

\cite{Carney79} normalized the ultraviolet excesses of 101 dwarfs by using the procedure of \cite{Sandage69} and calibrated them to the metal abundance $[Fe/H]$. This calibration could be used to evaluate metal abundances in the $UBV$ photometry.  \cite{Karaali03} improved this calibration by using a different procedure and a different set of $UBV$ data. Other works in different photometries followed the ones carried out in the $UBV$ for metallicity estimation. \cite{Buser90} calibrated the $[Fe/H]$ metal abundance to the normalized $\delta(U–-G)$ excess and $(G–-R)$ colour, simultaneously, in the $RGU$ photometry.  \cite{Stromgren66} defined the $m_1=(v–-b)–-(b–-y)$  colour difference as a metallicity indicator, where $v$, $b$, and $y$ are magnitudes for intermediate bands in his $uvby-\beta$ photometry and the $(B-–L)$ colour turned out to be a very sensitive metallicity index for F-–G spectral type stars in the $VBLUW$ photometry \citep{Walraven60, Trefzger95}. 

There are deviations between the calibrations obtained for the $UBV$ system. Fig. 15 of \cite{Buser92} compares these calibrations based on empirical data \citep{Carney79, Cameron85} or theoretical models \citep{Buser78, Buser85, Vandenberg85}. The reason for these differences originates from two sources: 1) Although researchers use the $UBV$ data of the same stars, the references and hence the $UBV$ magnitudes or colours therein may be different. 2) Different atmospheric parameters may be used by different researchers in estimation of the metallicities used for $[Fe/H]-\delta_{0.6}$ calibration \citep[cf.][]{Cayrel01}. 

The guillotine factors of \cite{Sandage69} are colour dependent, but not metallicity dependent. However, the isometallicity lines in the $(U–-B)–-(B–-V)$ two colour diagram are not parallel to each other for the whole colour range of $B–-V$ which indicates the dependence of the guillotine factors on the metallicity. This is the main topic of the paper. The data are presented in Section 2. The guillotine factors and the metallicity calibration are given in Section 3 and finally a short discussion is presented in Section 4.

\section{The data}

The PASTEL catalogue \citep{Soubiran10} is the main source of data for our study. 4259 stars with $4\leq \log g\leq5$ and with known metallicity and metallicity errors were selected as main sequence stars from the PASTEL catalogue. 3187 out of these stars that were not displaying a variability in their magnitudes which were tagged as ``star'', ``star in cluster'' and ``high proper motion stars'' in SIMBAD were used in the study. To obtain $UBV$ data, we consulted the specialized catalogues which are included in the General Catalogue of Photometric Data \citep{Hauck90}, which provided the data of 2073 stars.

To calibrate the metallicity more accurately, we selected 11 authors appearing in the PASTEL catalogue \citep{Soubiran10}, whose databases coincide the most with the 2073 stars in our study. These authors are: \citet{Valenti05}, \citet{Sousa08}, \citet{Ramirez05}, \citet{Santos04}, \citet{Fuhrmann08}, \citet{Luck06}, \citet{Mishenina04}, \citet{Nissen02}, \citet{Ryan03}, \citet{Spite1996} and \citet{Tomkin1999}. From all authors we collected a total of 701 stars of which \citet{Valenti05} determined the metal abundances of 472 stars. Hence, we reduced all the metallicities to \cite{Valenti05}'s using the calibrations between the metal abundances of common stars in the work of \citet{Valenti05} and other researchers. Table 2 gives the resulting star catalogue obtained by this procedure. The errors cited for the metal abundances belong to the original ones. The parallaxes were taken from the newly reduced Hipparcos catalogue \citep{vanLeeuwen07}. The $UBV$ data of stars in Table 2 have been dereddened by the following procedure \citep{Bahcall80}.

\begin{equation}
A_d(b)=A_{\infty}(b)\times(1-e^{\frac {-|d\sin (b)|}H})
\end{equation}
Here $b$ and $d$ are the Galactic latitude and the distance of the star (evaluated by means of its parallax), respectively. $H$ is the scaleheight for the interstellar dust which is adopted as 125 pc \citep{Marshall06} and $A_{\infty}(b)$ and $A_{d}(b)$ are the total absorptions for the model \citep{Schlegel98} and for the distance to the star respectively. $A_{\infty}(b)$ can be evaluated by means of Eq. 2
	
\begin{equation}
A_\infty(b)=3.1\times E(B-V),
\end{equation}
where $E_\infty(B-V)$ is the colour excess for the model taken from NASA Extragalactic Database\footnote{http://nedwww.ipac.caltech.edu/forms/calculator.html}. Then, $E_d(B-V)$, i.e. the colour excess for the corresponding star at the distance $d$ can be evaluated by Eq. 2 adopted for distance $d$ 

\begin{equation}
E_d(B-V)=A_d(b)/3.1,
\end{equation}
and can be used for the colour excess $E_d(U-B)$ evaluation:

\begin{equation}
E_d(U-B)=0.72 E_d(B-V)+0.05 E_d^2(B-V).
\end{equation}
Finally, the dereddened colour indices are:

\begin{eqnarray}
(B-V)_0=(B-V)-E_d(B-V)\\  \nonumber
(U-B)_0=(U-B)-E_d(U-B).
\end{eqnarray}         
The reduced ultraviolet excess $\delta_{0.6}$ is evaluated by the following equation which is obtained by the data of 133 stars with $0.575\leq(B-V)\leq0.625$ (Table 3):            

\begin{eqnarray}
\delta_{0.6}=-0.038(0.005)[Fe/H]^{3}-0.163(0.019)[Fe/H]^{2}\\ \nonumber
-0.302(0.017)[Fe/H]+0.012(0.004).
\end{eqnarray}
In this study, Karaali's guillotine factor is denoted by ``$f_K$'' and is calculated with $f_K=\delta_{0.6}/\delta$. 

%TABLE 2
%\begin{landscape}
\begin{table*}
\setlength{\tabcolsep}{4pt}  
\center
{\tiny
\caption{Data used for metallicity calibration. The columns give: CD, BD, HD or G (Giclas) number, ($\alpha$, $\delta$) and ($l$, $b$) equatorial and Galactic coordinates, distance (pc), dereddened $UBV$ data, reduced ultraviolet excess $\delta_{0.6}$, original metallicity $[Fe/H]$ and its error, metallicity reduced to \citet{Valenti05} system $[Fe/H]_{VF}$ and the author. The coordinates are defined as in ICRS.}
\begin{tabular}{cccccccccccccc}
\hline
Star & $\alpha$ & $\delta$ & $l$ & $b$ & $d$ & $V_{o}$ & $(B-V)_{o}$ & $(U-B)_{o}$ & $\delta_{0.6}$ & $[Fe/H]$ & $[Fe/H]_{err}$ & $[Fe/H]_{VF}$ & Author\\
\hline
HD000055   & 00 05 17.670 & -67 49 57.73 & 309.497 & -48.703 & 16 & 8.486 & 1.062 &  0.863 &  0.15 & -0.66 & 0.02 & -0.64 & Sousa \\
HD000101   & 00 05 54.739 & +18 14 05.83 & 108.005 & -43.313 & 37 & 7.431 & 0.554 &  0.036 &  0.08 & -0.28 & 0.04 & -0.25 & Ramirez \\
HD000142   & 00 06 19.215 & -49 04 30.76 & 321.587 & -66.387 & 26 & 5.694 & 0.514 &  0.021 & -0.03 &  0.10 & 0.03 &  0.14 & Valenti \\
HD000283   & 00 07 32.507 & -23 49 07.50 &~~48.982 & -79.560 & 33 & 8.675 & 0.795 &  0.336 &  0.13 & -0.55 & 0.03 & -0.52 & Valenti \\
HD000400   & 00 08 40.373 & +36 37 37.76 & 113.443 & -25.426 & 32 & 6.155 & 0.486 & -0.070 &  0.06 & -0.21 & 0.03 & -0.18 & Valenti \\
       ... &         ...  &         ...  &     ... &     ... &... &   ... &  ...  &   ...  &  ...  &  ...  & ...  &  ...  & ... \\
       ... &         ...  &         ...  &     ... &     ... &... &   ... &  ...  &   ...  &  ...  &  ...  & ...  &  ...  & ... \\
       ... &         ...  &         ...  &     ... &     ... &... &   ... &  ...  &   ...  &  ...  &  ...  & ...  &  ...  & ... \\
HD223498   & 23 50 05.768 & +02 52 38.04 &~~94.302 & -56.546 & 45 & 8.305 & 0.733 & 0.354  & -0.08 &  0.23 & 0.03 &  0.27 & Valenti \\
HD224022   & 23 54 38.598 & -40 18 00.16 & 341.042 & -72.356 & 28 & 6.013 & 0.572 & 0.106  & -0.05 &  0.15 & 0.06 &  0.19 & Sousa \\
HD224156   & 23 55 32.411 & +03 30 04.95 &~~97.015 & -56.531 & 30 & 7.685 & 0.746 & 0.346  &  0.01 & -0.03 & 0.03 &  0.00 & Valenti \\
HD224383   & 23 57 33.478 & -09 38 50.59 &~~84.366 & -68.386 & 48 & 7.826 & 0.629 & 0.137  &  0.02 & -0.04 & 0.03 & -0.01 & Valenti \\
HD224619   & 23 59 28.388 & -20 02 05.06 &~~60.981 & -76.155 & 25 & 7.456 & 0.740 & 0.281  &  0.06 & -0.20 & 0.01 & -0.17 & Sousa \\
\hline
\end{tabular}  
}
\end{table*}

\section {Methods}
\subsection{New Guillotine Factors}

\citet{Sandage69} estimated guillotine factors without considering the effect of  metallicity. However, Fig. 2 shows that the colour gradients for any two different isometallicity lines are not equal to each other, i.e. $\frac {|AB|}{|DE|}\neq \frac{|AC|}{|DF|}$, which indicates the dependence of guillotine factors on metallicity. Additionally, the $(U-B)_M$ colours in Table 1 correspond to the stars with less metallicity than the Hyades cluster. But, the metallicity gradients for the isometallicity line with $[Fe/H]=0.5$ dex in Fig. 2 are rather different than the ones for relatively metal poor stars which indicates that guillotine factors for metal rich stars should be different than those of metal poor stars of the same $B-V$ colour index.

%FIGURE 2
\begin{figure}
\begin{center}
\includegraphics[scale=0.375, angle=0]{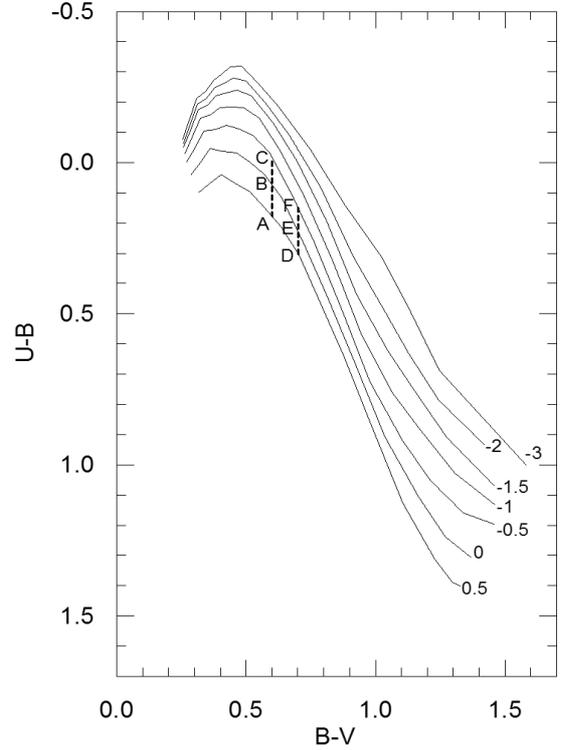}
\caption[] {Synthetic isometallicity lines for $UBV$ photometry taken from a stellar model of \citet{Lejeune97}.}
\end{center}
\end{figure}

%FIGURE 3
\begin{figure}
\begin{center}
\includegraphics[scale=0.37, angle=0]{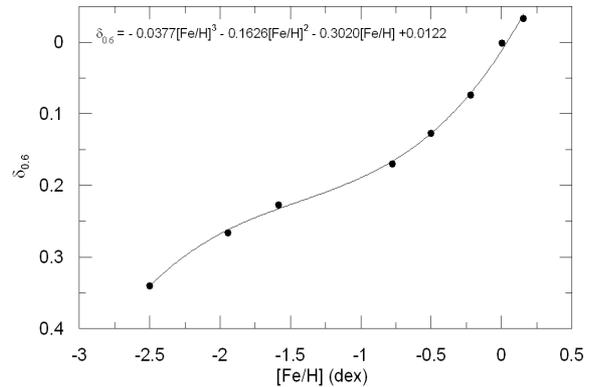}
\caption[] {Metallicity versus ultraviolet excess calibration for 133 stars with $0.575\leq B-V\leq0.625$.}
\end{center}
\end{figure}

%TABLE 3
\begin{table*}
\setlength{\tabcolsep}{.75pt}  
\center
{\scriptsize
\caption{Stars with $0.575\leq B-V\leq0.625$ used for calibration of $\delta_{0.6}$ to metallicity. Symbols are the same with Table 2.}
\begin{tabular}{ccccccccccccccc}
\hline
Star & $(B-V)_o$ & $(U-B)_o$ & $[Fe/H]$ & $\delta_{(0.6)}$ & Star & $(B-V)_o$ & $(U-B)_o$ & $[Fe/H]$ & $\delta_{(0.6)}$ & Star & $(B-V)_o$ & $(U-B)_o$ & $[Fe/H]$ & $\delta_{(0.6)}$\\
\hline
HD020407 & 0.576 & -0.023 & -0.42 & 0.128 & HD155918 & 0.592 & -0.017 & -0.64 & 0.139 & HD221146 & 0.611 & 0.214 & 0.11 & -0.070 \\
HD043745 & 0.577 & 0.078 & 0.13 & 0.028 & HD024040 & 0.593 & 0.198 & 0.21 & -0.074 & HD010519 & 0.612 & 0.027 & -0.58 & 0.118 \\
HD003149 & 0.578 & 0.061 & -0.07 & 0.046 & HD039091 & 0.593 & 0.094 & 0.05 & 0.030 & HD013043 & 0.612 & 0.145 & 0.09 & 0.000 \\
HD134088 & 0.578 & -0.059 & -0.75 & 0.166 & HD102158 & 0.593 & 0.038 & -0.47 & 0.086 & HD033021 & 0.612 & 0.080 & -0.14 & 0.065 \\
HD104800 & 0.578 & -0.059 & -0.82 & 0.166 & HD196800 & 0.593 & 0.137 & 0.16 & -0.013 & HD084737 & 0.612 & 0.143 & 0.17 & 0.002 \\
HD145809 & 0.578 & 0.057 & -0.24 & 0.050 & HD221830 & 0.593 & 0.057 & -0.40 & 0.067 & HD054351 & 0.613 & 0.115 & -0.05 & 0.031 \\
HD188510 & 0.578 & -0.131 & -1.64 & 0.238 & HD061383 & 0.594 & 0.024 & -0.48 & 0.101 & HD064090 & 0.614 & -0.129 & -1.80 & 0.276 \\
BD+660268 & 0.580 & -0.147 & -2.09 & 0.257 & BD+730943 & 0.594 & 0.016 & -0.37 & 0.109 & HD030562 & 0.614 & 0.181 & 0.26 & -0.034 \\
HD286891 & 0.580 & -0.040 & -0.56 & 0.150 & HD070110 & 0.594 & 0.142 & 0.15 & -0.017 & HD068978 & 0.614 & 0.067 & 0.02 & 0.080 \\
HD030649 & 0.580 & 0.029 & -0.49 & 0.081 & HD088986 & 0.594 & 0.156 & 0.09 & -0.031 & HD139324 & 0.614 & 0.156 & 0.15 & -0.009 \\
HD078366 & 0.580 & 0.039 & 0.08 & 0.071 & HD096700 & 0.594 & 0.065 & -0.18 & 0.060 & HD094835 & 0.615 & 0.116 & 0.13 & 0.032 \\
HD126681 & 0.581 & -0.115 & -1.16 & 0.226 & HD009782 & 0.595 & 0.086 & 0.09 & 0.040 & HD178496 & 0.615 & 0.119 & -0.26 & 0.029 \\
HD029461 & 0.581 & 0.145 & 0.25 & -0.034 & HD009782 & 0.595 & 0.086 & 0.15 & 0.040 & HD120066 & 0.615 & 0.147 & 0.11 & 0.001 \\
HD006500 & 0.582 & -0.009 & -0.60 & 0.121 & HD045391 & 0.595 & 0.016 & -0.50 & 0.110 & HD166435 & 0.615 & 0.086 & 0.04 & 0.062 \\
HD007983 & 0.582 & -0.028 & -0.60 & 0.140 & HD073524 & 0.595 & 0.116 & 0.12 & 0.010 & BD+591609 & 0.616 & 0.070 & -0.45 & 0.079 \\
HD115383 & 0.582 & 0.093 & 0.28 & 0.019 & HD067458 & 0.596 & 0.036 & -0.19 & 0.091 & HD003074 & 0.616 & 0.146 & 0.00 & 0.003 \\
HD196850 & 0.582 & 0.071 & -0.11 & 0.041 & HD110898 & 0.597 & 0.009 & -0.38 & 0.119 & HD066171 & 0.616 & 0.057 & -0.31 & 0.092 \\
HD199288 & 0.582 & -0.045 & -0.63 & 0.157 & HD004307 & 0.597 & 0.069 & -0.19 & 0.059 & HD208704 & 0.617 & 0.109 & -0.08 & 0.041 \\
HD165499 & 0.583 & 0.055 & 0.01 & 0.058 & HD020807 & 0.598 & 0.001 & -0.26 & 0.128 & HD152792 & 0.617 & 0.091 & -0.31 & 0.059 \\
HD044120 & 0.583 & 0.080 & 0.10 & 0.033 & HD143761 & 0.598 & 0.083 & -0.20 & 0.046 & HD250792 & 0.618 & -0.043 & -1.07 & 0.195 \\
HD133161 & 0.583 & 0.165 & 0.21 & -0.052 & HD150433 & 0.599 & 0.047 & -0.38 & 0.083 & HD001832 & 0.618 & 0.121 & -0.03 & 0.031 \\
HD158226 & 0.584 & -0.028 & -0.52 & 0.142 & HD070923 & 0.600 & 0.097 & 0.12 & 0.034 & HD008262 & 0.618 & 0.091 & -0.16 & 0.061 \\
BD+592407 & 0.584 & -0.103 & -1.95 & 0.217 & HIP043595 & 0.600 & -0.040 & -0.80 & 0.171 & HD051419 & 0.618 & 0.069 & -0.40 & 0.083 \\
HD171990 & 0.585 & 0.114 & 0.07 & 0.001 & HD006434 & 0.600 & -0.012 & -0.52 & 0.143 & HD053705 & 0.618 & 0.045 & -0.21 & 0.107 \\
HD083529 & 0.585 & 0.019 & -0.25 & 0.096 & HD150706 & 0.600 & 0.076 & -0.01 & 0.055 & HD097998 & 0.619 & 0.052 & -0.41 & 0.101 \\
HD206332 & 0.585 & 0.169 & 0.27 & -0.054 & HD088218 & 0.600 & 0.148 & -0.14 & -0.017 & HD216435 & 0.619 & 0.169 & 0.24 & -0.016 \\
HD059360 & 0.586 & 0.087 & -0.12 & 0.029 & HD088725 & 0.600 & -0.016 & -0.70 & 0.147 & HD036283 & 0.620 & 0.083 & -0.31 & 0.071 \\
HD129290 & 0.586 & 0.062 & -0.13 & 0.054 & HD107146 & 0.600 & 0.071 & -0.03 & 0.060 & HD031966 & 0.621 & 0.215 & 0.13 & -0.060 \\
HD018709 & 0.586 & -0.003 & -0.26 & 0.119 & HD090508 & 0.601 & 0.054 & -0.30 & 0.078 & HD038858 & 0.621 & 0.086 & -0.23 & 0.069 \\
HD131117 & 0.586 & 0.087 & 0.14 & 0.029 & HD165401 & 0.601 & 0.021 & -0.36 & 0.111 & HD034411 & 0.622 & 0.121 & 0.12 & 0.035 \\
HD170778 & 0.586 & 0.087 & 0.00 & 0.029 & HD222033 & 0.601 & 0.121 & 0.19 & 0.011 & HD071148 & 0.622 & 0.118 & 0.02 & 0.038 \\
HD014056 & 0.587 & -0.012 & -0.61 & 0.129 & HD010226 & 0.602 & 0.134 & 0.22 & 0.000 & HD071881 & 0.622 & 0.134 & -0.05 & 0.022 \\
HD111367 & 0.587 & 0.086 & -0.18 & 0.031 & HD211415 & 0.603 & 0.059 & -0.20 & 0.076 & HD183658 & 0.623 & 0.158 & 0.05 & -0.001 \\
G 232-18 & 0.587 & -0.019 & -0.80 & 0.136 & HD218209 & 0.604 & 0.049 & -0.46 & 0.087 & HD042618 & 0.624 & 0.119 & -0.11 & 0.040 \\
HD124553 & 0.587 & 0.143 & 0.28 & -0.026 & HD056274 & 0.605 & -0.018 & -0.55 & 0.155 & HD088371 & 0.624 & 0.109 & -0.31 & 0.050 \\
HD093745 & 0.587 & 0.111 & 0.08 & 0.006 & HD073668 & 0.605 & 0.079 & 0.00 & 0.058 & HD199960 & 0.624 & 0.213 & 0.27 & -0.054 \\
HD041330 & 0.588 & 0.051 & -0.14 & 0.067 & HD095128 & 0.605 & 0.123 & 0.04 & 0.014 & HD018757 & 0.625 & 0.128 & -0.28 & 0.032 \\
HD019373 & 0.588 & 0.113 & 0.16 & 0.005 & HD164427 & 0.605 & 0.117 & 0.13 & 0.020 & CD-2808426 & 0.625 & 0.019 & -0.64 & 0.141 \\
HD052711 & 0.588 & 0.058 & -0.10 & 0.060 & HD197076 & 0.607 & 0.071 & -0.09 & 0.068 & HD147231 & 0.625 & 0.216 & 0.00 & -0.056 \\
HD168871 & 0.588 & 0.046 & -0.09 & 0.072 & HD009224 & 0.608 & 0.108 & 0.00 & 0.032 & HD179140 & 0.625 & 0.162 & 0.12 & -0.002 \\
HD088742 & 0.589 & 0.073 & -0.05 & 0.046 & HD149612 & 0.608 & 0.024 & -0.45 & 0.116 & HD196068 & 0.625 & 0.273 & 0.31 & -0.113 \\
HD283807 & 0.590 & 0.000 & -0.58 & 0.120 & HD114729 & 0.609 & 0.044 & -0.26 & 0.097 & HD200565 & 0.625 & 0.112 & -0.06 & 0.048 \\
HD121004 & 0.591 & -0.056 & -0.80 & 0.177 & HD118475 & 0.609 & 0.166 & 0.10 & -0.025 & BD+38 4955 & 0.660 & -0.160 & -2.50 & 0.340 \\
HD001388 & 0.591 & 0.088 & 0.00 & 0.033 & HD223238 & 0.609 & 0.128 & 0.02 & 0.013 &  &  &  &  &  \\
HD016623 & 0.591 & 0.007 & -0.45 & 0.114 & HD134060 & 0.610 & 0.139 & 0.09 & 0.003 &  &  &  &  &  \\
\hline
\end{tabular}
}  
\end{table*}

%FIGURE 4
\begin{figure*}
\begin{center}
\includegraphics[scale=0.65, angle=0]{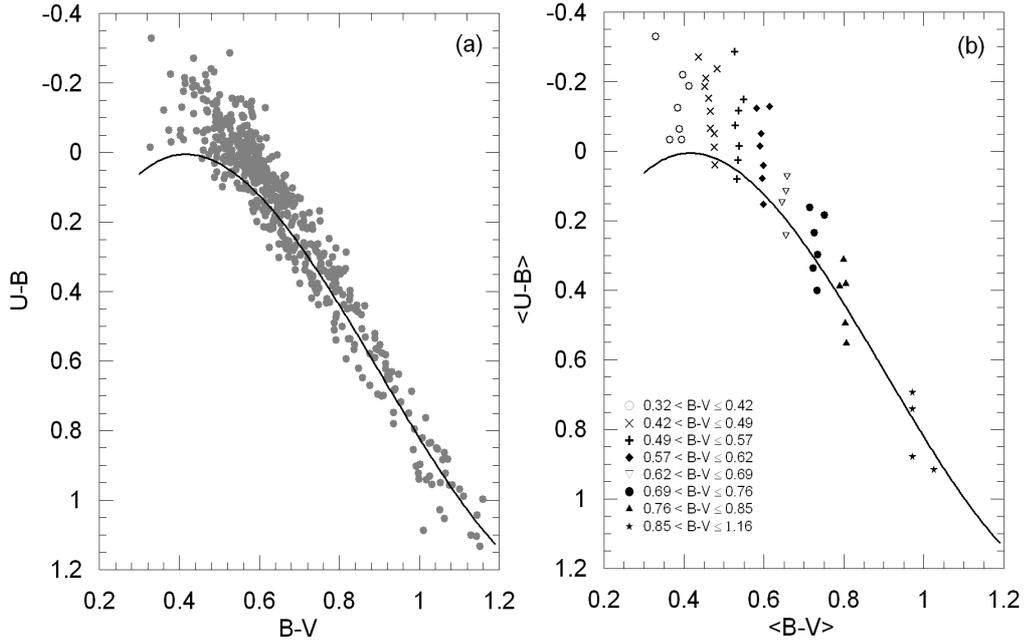}
\caption[] {$(U-B)-–(B-V)$ two colour diagram for the whole sample (a) and for 50 bins of eight subsamples in Table 4 (b).}
\end{center}
\end{figure*}  

%FIGURE 5
\begin{figure*}
\begin{center}
\includegraphics[scale=0.7, angle=0]{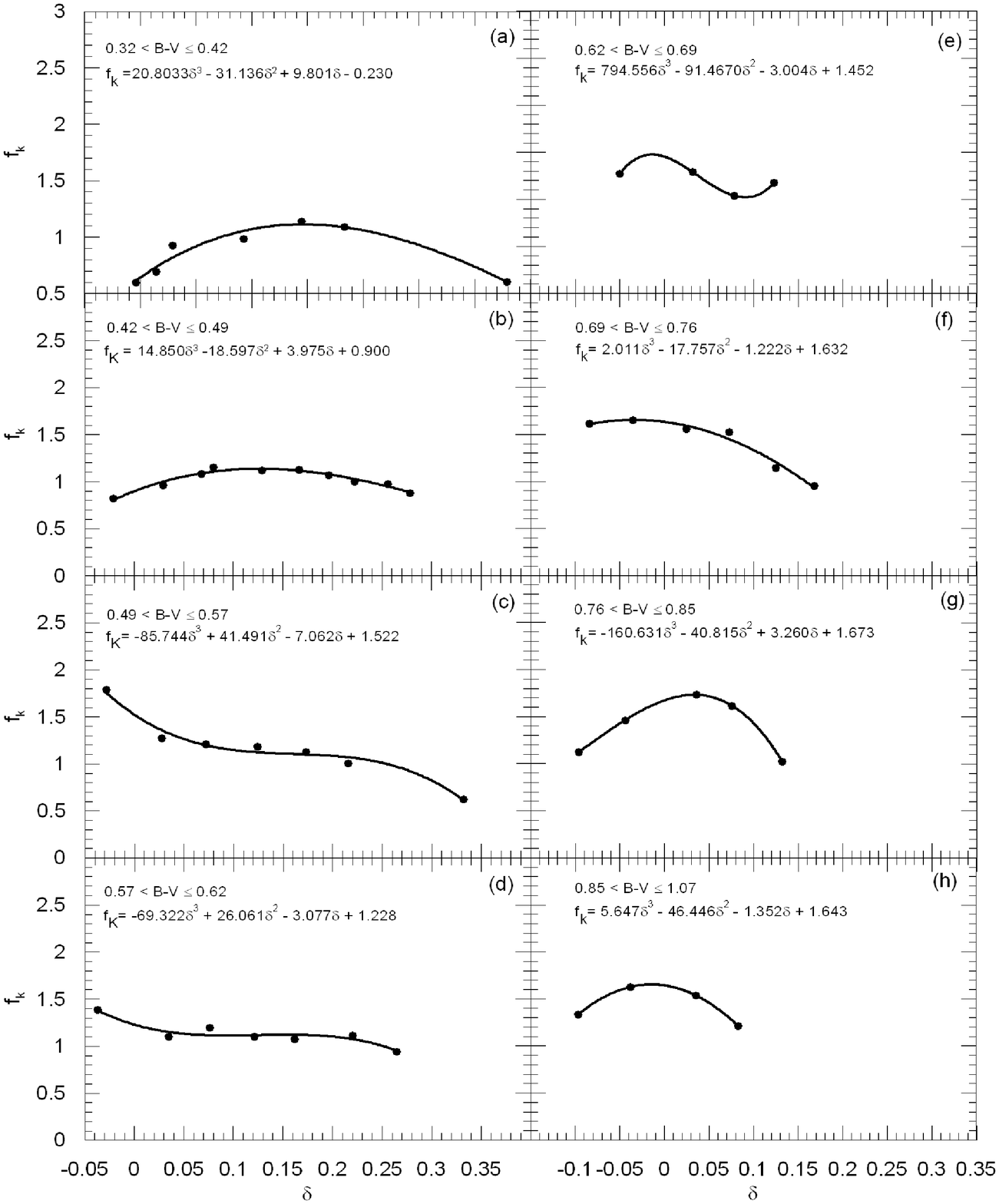}
\caption[] {Calibration of ultraviolet excess ($\delta$) to the guillotine factor $f_K$ for eight subsamples. }
\end{center}
\end{figure*}

Then we decided that it should be more appropriate and useful to estimate metallicity dependent guillotine factors. First, we used 133 stars with colour index $0.575\leq B-V \leq0.625$ and calibrated their ultraviolet excess to the metallicity. The calibration (Eq. 6) provides ultraviolet excess reduced to $B-V=0.60$ for any star with metal abundance $[Fe/H]\geq-2.5$ dex. Thus, we can use the calibration obtained from Fig. 3 to estimate reduced $\delta_{0.6}$ ultraviolet excess for stars with metal abundance $-2.5\leq[Fe/H]\leq0.15$ dex. 

Next, we separated the stars in Table 2 into eight subsamples with colour indices $0.32<B-V\leq0.42$, $0.42<B-V\leq 0.49$, $0.49<B-V\leq0.57$, $0.57<B-V\leq0.62$, $0.62<B-V\leq0.69$, $0.69<B-V\leq0.76$, $0.76<B-V\leq0.85$, $0.85<B-V\leq1.16$ and obtained calibrations for the guillotine factors as explained in the following: The number of these colour intervals and their ranges had been decided such as to obtain a constant metallicity gradient for each $B-V$ interval. For example, the ranges for bluer stars, where the metallicity gradient is relatively large were adopted smaller than for the colour interval, $0.85<B-V\leq1.16$, where the metallicity gradient is rather smooth. The $(U-B)-(B-V)$ colour diagrams of the whole sample and eight subsamples are shown in Fig. 4.

Each subsample was divided into bins and mean $\delta$, $B-V$, $U–-B$, $[Fe/H]$, $(U-B)_H$, $\delta_{0.6}$ and $f_K$ values were evaluated for each subsample (Table 4). A total of 532 stars could be used in the calibration of guillotine factors. Ultraviolet–-excess $\delta$ of a sample star whose $U-B$ colour index is close to that of a Hyades star of the same $B-V$ colour index is rather small. Hence, $f_K$=$\delta_{0.6}$/$\delta$ becomes rather large for such stars and they are not reliable. These abnormal $f_K$ values may be as large as 20, for example. Also, we noted that the $f_K$ values of some stars were negative. The reason of these unreliable values originate from the errors in $U-B$ colour index. After rejection the stars with large and negative $f_K$ values the number of stars reduced from 701 to 532. The numbers of stars used in each bin is given in the last column of Table 4.  

The calibration of $\delta$ to $f_K$ is given in Fig. 5. One notices that there is a smooth relation in all panels, and the trend of the guillotine factor varies in different panels. In panels (a), (b), (g) and (h), $f_K$ assumes its maximum at the intermediate values of $\delta$, whereas in panels (c), (d), (e) and (f)  the maximum of $f_K$ corresponds to the negative values of $\delta$, i.e. metal rich stars. Fig. 6 shows the calibration of $B-V$ to $f_K$ for three ultraviolet excess, i.e. $\delta=-0.05$, +0.05 and +0.15, just to show that one can obtain continuous transitions between colour index and ultraviolet excess. Since Fig. 5 is divided according to colours and the fits are in better agreement with data, the obtained equations will be more precise. Therefore, we prefer the equations obtained from Fig. 5.

Fig. 7 shows the distribution of the guillotine factors as a function of metallicity. The lower limit for the guillotine factors of \cite{Sandage69} is $f_S=1$ (panel b), corresponding to the colour index $B–-V=0.60$, whereas the one estimated in this work which may be less than 0.5, which is not colour dependent, but corresponds to metal abundance $[Fe/H]\approx0$ dex. 

%TABLE 4
\begin{table*}
\center
\caption{Ultraviolet excess $\delta$, reduced ultraviolet excess $\delta_{0.6}$ and new guillotine factors $f_{K}$ for each bin of eight sub–-samples. The symbols are explained in the text.} 
\begin{tabular}{cccccccc}
\hline
$\langle B-V\rangle$ & $\langle U-B\rangle$ & $\langle [Fe/H] \rangle$ & $\langle (U-B)_{H} \rangle$ & $\delta$ & $\delta_{0.6}$ & $f_{K}$ & N \\
\hline
\multicolumn{8}{c}{$0.32<B-V\leq0.42$}\\
0.394 &     -0.034 &     -0.053 &      0.012 &      0.046 &      0.028 &      0.595 &          2 \\
0.364 &     -0.033 &     -0.110 &      0.031 &      0.064 &      0.043 &      0.695 &          2 \\
0.388 &     -0.064 &     -0.224 &      0.014 &      0.079 &      0.071 &      0.924 &          2 \\
0.384 &     -0.126 &     -0.584 &      0.017 &      0.143 &      0.140 &      0.985 &          2 \\
0.412 &     -0.188 &     -1.424 &      0.008 &      0.195 &      0.221 &      1.135 &          2 \\
0.396 &     -0.220 &     -1.836 &      0.012 &      0.233 &      0.254 &      1.092 &          2 \\
0.329 &     -0.330 &     -1.530 &      0.050 &      0.379 &      0.229 &      0.603 &          1 \\
\multicolumn{8}{c}{$0.42<B-V\leq0.49$}\\
0.477 &      0.039 &      0.082 &      0.018 &     -0.021 &     -0.014 &      0.819 &          3 \\
0.475 &     -0.012 &     -0.054 &      0.017 &      0.029 &      0.027 &      0.964 &          6 \\
0.475 &     -0.051 &     -0.228 &      0.017 &      0.068 &      0.073 &      1.082 &          2 \\
0.466 &     -0.066 &     -0.307 &      0.014 &      0.080 &      0.090 &      1.152 &          4 \\
0.466 &     -0.115 &     -0.633 &      0.014 &      0.129 &      0.145 &      1.119 &          4 \\
0.462 &     -0.153 &     -0.981 &      0.013 &      0.166 &      0.187 &      1.128 &          4 \\
0.452 &     -0.186 &     -1.268 &      0.010 &      0.196 &      0.210 &      1.072 &          6 \\
0.455 &     -0.211 &     -1.443 &      0.011 &      0.222 &      0.223 &      1.003 &          3 \\
0.482 &     -0.237 &     -1.763 &      0.019 &      0.256 &      0.249 &      0.973 &          2 \\
0.436 &     -0.271 &     -1.745 &      0.007 &      0.278 &      0.244 &      0.879 &          1 \\
\multicolumn{8}{c}{$0.49<B-V\leq0.57$}\\
0.532 &      0.080 &      0.188 &      0.052 &     -0.028 &     -0.053 &      1.787 &         12 \\
0.535 &      0.026 &     -0.085 &      0.054 &      0.028 &      0.036 &      1.273 &         24 \\
0.537 &     -0.016 &     -0.307 &      0.056 &      0.072 &      0.087 &      1.209 &         38 \\
0.528 &     -0.075 &     -0.641 &      0.049 &      0.124 &      0.146 &      1.182 &         27 \\
0.536 &     -0.117 &     -1.069 &      0.056 &      0.173 &      0.194 &      1.127 &         10 \\
0.549 &     -0.150 &     -1.414 &      0.065 &      0.216 &      0.217 &      1.006 &          4 \\
0.526 &     -0.286 &     -1.211 &      0.046 &      0.332 &      0.206 &      0.623 &          1 \\
\multicolumn{8}{c}{$0.57<B-V\leq0.62$}\\
0.598 &      0.153 &      0.161 &      0.116 &     -0.037 &     -0.042 &      1.387 &         19 \\
0.595 &      0.078 &     -0.096 &      0.113 &      0.035 &      0.039 &      1.098 &         23 \\
0.598 &      0.041 &     -0.321 &      0.117 &      0.076 &      0.089 &      1.195 &         29 \\
0.590 &     -0.015 &     -0.546 &      0.107 &      0.121 &      0.133 &      1.097 &         21 \\
0.593 &     -0.051 &     -0.850 &      0.111 &      0.162 &      0.174 &      1.073 &          5 \\
0.581 &     -0.124 &     -1.710 &      0.096 &      0.220 &      0.245 &      1.112 &          4 \\
0.614 &     -0.129 &     -1.801 &      0.136 &      0.265 &      0.249 &      0.941 &          1 \\
\multicolumn{8}{c}{$0.62<B-V\leq0.69$}\\
0.655 &      0.243 &      0.201 &      0.192 &     -0.050 &     -0.057 &      1.274 &         34 \\
0.645 &      0.147 &     -0.111 &      0.179 &      0.032 &      0.042 &      1.288 &         22 \\
0.655 &      0.115 &     -0.276 &      0.193 &      0.078 &      0.081 &      1.038 &         29 \\
0.658 &      0.073 &     -0.611 &      0.196 &      0.123 &      0.143 &      1.177 &         12 \\
\multicolumn{8}{c}{$0.69<B-V\leq0.76$}\\
0.732 &      0.401 &      0.387 &      0.317 &     -0.084 &     -0.133 &      1.611 &          6 \\
0.722 &      0.335 &      0.218 &      0.300 &     -0.035 &     -0.064 &      1.650 &         13 \\
0.734 &      0.296 &     -0.098 &      0.321 &      0.025 &      0.038 &      1.555 &         11 \\
0.726 &      0.234 &     -0.429 &      0.306 &      0.073 &      0.109 &      1.525 &         17 \\
0.714 &      0.161 &     -0.604 &      0.286 &      0.125 &      0.141 &      1.143 &          4 \\
0.751 &      0.183 &     -0.886 &      0.351 &      0.168 &      0.164 &      0.955 &          2 \\
\multicolumn{8}{c}{$0.76<B-V\leq0.85$}\\
0.806 &      0.550 &      0.328 &      0.454 &     -0.096 &     -0.106 &      1.127 &          5 \\
0.803 &      0.491 &      0.220 &      0.447 &     -0.044 &     -0.063 &      1.461 &         10 \\
0.789 &      0.385 &     -0.175 &      0.422 &      0.036 &      0.060 &      1.737 &         11 \\
0.805 &      0.378 &     -0.458 &      0.453 &      0.075 &      0.117 &      1.614 &         17 \\
0.799 &      0.308 &     -0.583 &      0.440 &      0.132 &      0.134 &      1.024 &          8 \\
\multicolumn{8}{c}{$0.85<B-V\leq1.16$}\\
0.972 &      0.878 &      0.360 &      0.782 &     -0.097 &     -0.121 &      1.332 &         11 \\
1.025 &      0.915 &      0.215 &      0.877 &     -0.038 &     -0.062 &      1.627 &         11 \\
0.971 &      0.741 &     -0.165 &      0.776 &      0.035 &      0.055 &      1.539 &         21 \\
0.971 &      0.693 &     -0.373 &      0.777 &      0.083 &      0.100 &      1.214 &         22 \\
\hline
\end{tabular}
\end{table*}  

%FIGURE 6
\begin{figure}
\begin{center}
\includegraphics[scale=0.37, angle=0]{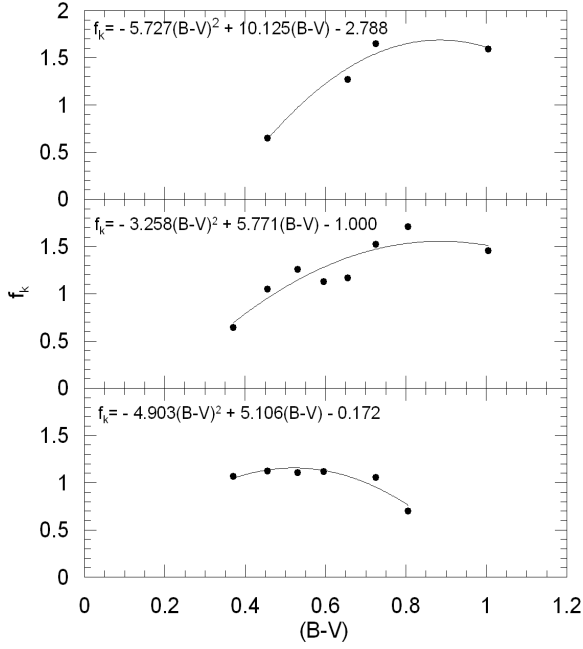}
\caption[] {Calibration of $B-V$ colour to the guillotine factor $f_K$ for three ultraviolet excesses, $\delta=-0.05$, +0.05, and +0.15.}
\end{center}
\end{figure}

%FIGURE 7
\begin{figure}
\begin{center}
\includegraphics[scale=0.35, angle=0]{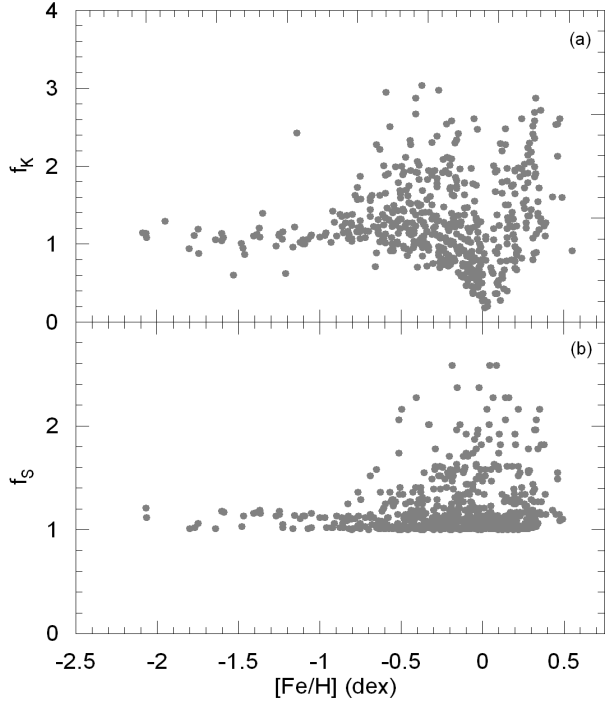}
\caption[] {Guillotine factors versus metallicity: (a) for metallicity dependent guillotine factors ($f_{K}$); (b) for guillotine factors free of metallicity given by \citet{Sandage69}($f_{S}$).}
\end{center}
\end{figure}

%FIGURE 8
\begin{figure}
\begin{center}
\includegraphics[scale=0.35, angle=0]{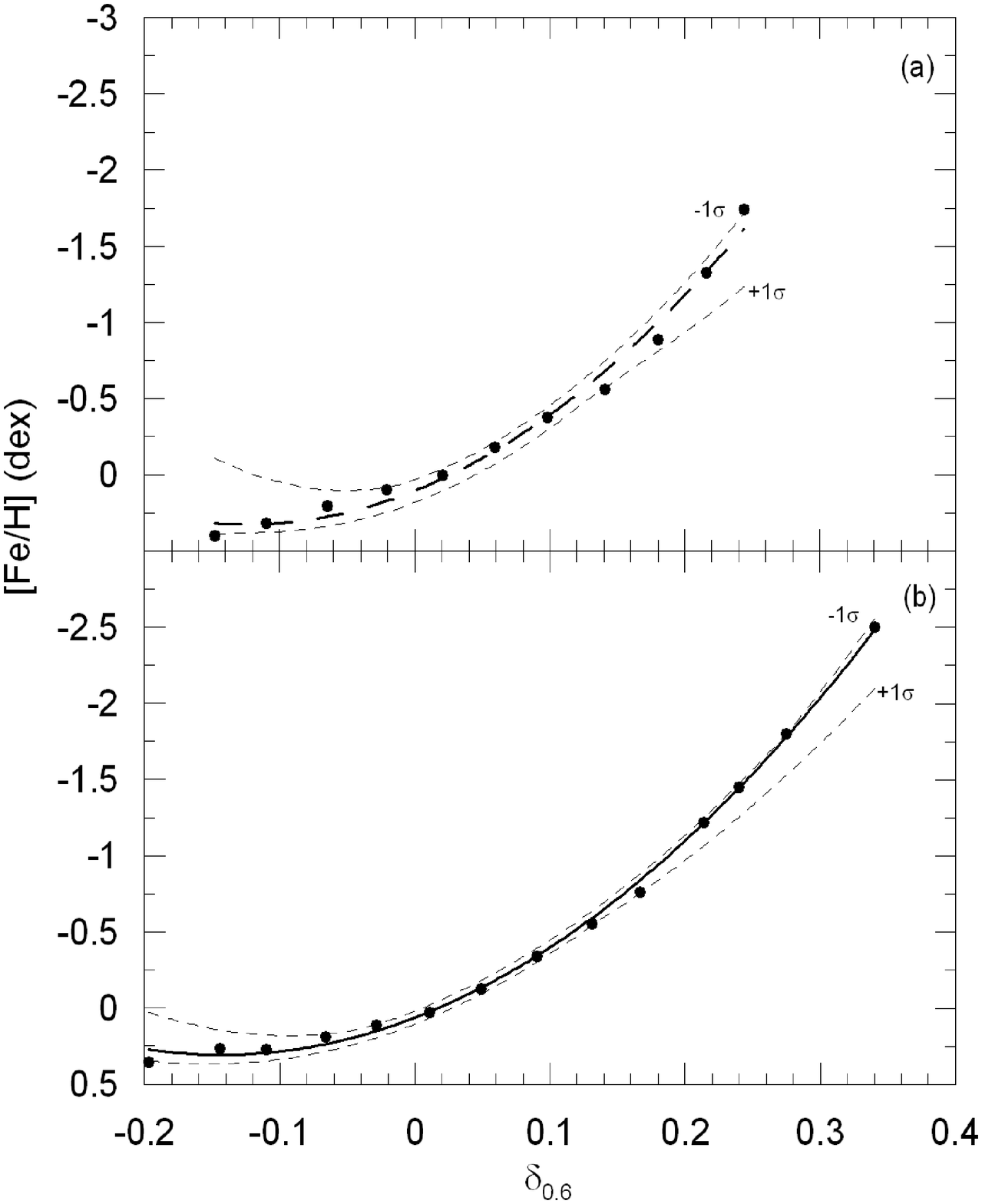}
\caption[] {Metallicity calibration based on the metallicity dependent guillotine factors (a) and the ones adopted from \citet{Sandage69} (b). The dashed lines denote the $\pm1\sigma$ prediction levels.}
\end{center}
\end{figure}

\subsection{New Metallicity Calibration}

We used the calibrations in Fig. 5 and assigned guillotine factors, $f_K$, for 701 stars with metallicity $-1.76\leq[Fe/H]\leq+0.40$ dex. The combinations of $f_K$ and the ultraviolet excess $\delta$ gives the reduced ultraviolet excess for each star, i.e. $\delta_{0.6}=f_K\times \delta$. Then we divided the interval $-0.15\leq\delta_{0.6} \leq0.24$ into 11 scans and adopted the centroid of each scan as a locus point to fit a second order polynomial (Fig. 8) to the couple ($\delta_{0.6}$, $[Fe/H]$). The full equation of the polynomial is

\begin{eqnarray}
[Fe/H]=-14.316(1.919)\delta_{0.6}^2+3.557(0.285)\delta_{0.6}\\ \nonumber
+0.105(0.039).
\end{eqnarray}
Eq. 7 provides metallicities by means of new guillotine factors. We used the guillotine factors of \cite{Sandage69} and evaluated another set of reduced $\delta_{0.6}$ ultraviolet excess for the same star sample. Their fit to the corresponding metallicities is given in Fig. 8 and in the following formula:

\begin{eqnarray}
[Fe/H]=-11.612(0.496)\delta_{0.6}^2+3.419(0.100)\delta_{0.6}\\ \nonumber
+0.057(0.017).
\end{eqnarray}

\subsection{Application of the Method}

Now, we have two metallicity calibrations, based on the new guillotine factors estimated in this work and on the guillotine factors adopted from \cite{Sandage69}. We applied these calibrations to two sets of  data with $-1.76\leq[Fe/H]\leq0.4$ dex taken from \cite{Karaali03} and \cite{Karatas06} and we compared the evaluated metallicities with the original ones for two calibrations. The metallicities of 75 stars in the first set were estimated spectroscopically, whereas those for 469 stars are based on photometry. Stars in two sets just mentioned and the sample stars do not overlap. The results for the data of \cite{Karaali03} are given in Table 5. The clarification of the symbols is as follows: Hip No: the Hipparcos number, $\delta$ the ultraviolet excess, $f_K$ and $f_S$: the guillotine factors estimated in this work and adopted from \cite{Sandage69}, respectively, $\delta_{0.6}(K)$ and $\delta_{0.6}(S)$: ultraviolet excess reduced by means of $f_K$ and $f_S$, respectively, $[Fe/H]_{obs}$, $[Fe/H]_K$ and $[Fe/H]_S$: original metallicities taken from the literature and metallicities evaluated via Eqs. 7 and 8, respectively, $\Delta[Fe/H]_K$ and $\Delta[Fe/H]_S$: residuals for two calibrations, where $K$ and $S$ refer to the data evaluated by means of the guillotine factors estimated in this work and the ones adopted from \cite{Sandage69}. The results for 469 stars are not given here in order to avoid space consuming. However, their statistics are given in Table 6 together with the ones of 75 stars in the first set.

Comparison of the mean and standard deviations for the residuals of two calibrations for four metallicity intervals, i.e. $-1.76<[Fe/H]\leq-1$, $-1<[Fe/H]\leq-0.5$, $-0.5<[Fe/H]\leq0$ and $0<[Fe/H]\leq+0.4$ dex shows that there are statistical differences between two calibrations. In Table 6a, where the statistics for the first set (75 stars) is presented, the agreement is only for the interval $-1<[Fe/H]\leq-0.5$ dex, whereas for other metallicity intervals, the calibration based on metallicity dependent guillotine factors $f_K$ favorites. The largest differences between two sets of statistics belong to the metal poor stars, i.e. $-1.76<[Fe/H]\leq-1$ dex. In Table 6b, where the statistics correspond to a larger set of data (469 stars) and where the metallicities were determined photometrically, the agreement between two calibrations is only for the metallicity interval $-1.76<[Fe/H]\leq-1$ dex. The mean deviation of the residuals estimated via guillotine factors $f_S$, for the metallicity interval $-0.5<[Fe/H]\leq0$ dex, is a bit smaller than the ones estimated via $f_K$ (0.01 and -0.03 respectively), whereas, for two metallicity intervals, i.e. $-1<[Fe/H]\leq-0.5$ and $0<[Fe/H]\leq+0.4$ dex, the mean deviations corresponding to $f_K$ are much smaller than the ones of $f_S$.    

The comparison of the residuals for all metallicities, i.e. $-1.76<[Fe/H]\leq+0.4$ dex, estimated by means of two calibrations (Fig. 9 and Fig. 10) confirms the advantage of the calibration based on metallicity dependent guillotine factors. There is a small correlation for the residuals in the lower panel in Fig. 9, $R^2=0.25$, which corresponds to guillotine factors $f_S$, whereas in the upper panel where the residuals were based on the guillotine factors $f_K$, the distribution of the points about the line of zero residual is almost homogeneous resulting a zero correlation coefficient, $R^2=0.00$.   

In Fig. 10, the residuals are calibrated to linear equations of the metallicities. The panels (a) and (b) correspond to the residuals estimated via metallicity dependent guillotine factors $f_K$ and those to the metal free ones $f_S$. The inclinations of the lines are 0.10 and 0.20 for panels (a) and (b), respectively, favoring the $f_K$ factors. Also, the correlation coefficients i.e. $R^2=0.03$ and $R^2=0.11$, for panels (a) and (b) respectively, confirm our argument. That is, by less correlation coefficient, we assume a relatively homogeneous distribution for the residuals in panel (a).

It is interesting that there are small differences in statistics for two sets of data which can be explained either by $UBV$ data or metallicities used. We should remind that metallicities for the first set (75 stars) were estimated spectroscopically, whereas for the second set (469 stars) a photometric procedure was used.                         

%TABLE 5
\begin{table*}
\setlength{\tabcolsep}{2pt}
\center
{\scriptsize
\caption{Comparison of the original metallicities taken from the literature with the ones evaluated by using two different calibrations (Eqs. 7 and 8). The symbols are explained in the text.}
\begin{tabular}{lcccccccccc}
\hline
Hip No & $\delta$ & $f_{K}$ & $f_{S}$ & $(\delta_{0.6})_{K}$ & $(\delta_{0.6})_{S}$ & $[Fe/H]_{obs}$ & $[Fe/H]_{K}$ & $[Fe/H]_{S}$ & $\Delta[Fe/H]_{K}$ & $\Delta[Fe/H]_{S}$\\
\hline
      1599 &      0.085 &      1.112 &       1.01 &      0.095 &      0.086 &      -0.26 &     -0.362 &     -0.325 &      0.102 &      0.065 \\
      3206 &      0.020 &      1.597 &       1.61 &      0.032 &      0.032 &      -0.06 &     -0.024 &     -0.066 &     -0.036 &      0.006 \\
      6702 &      0.021 &      0.976 &       1.14 &      0.021 &      0.024 &       0.16 &      0.025 &     -0.033 &      0.135 &      0.193 \\
      8102 &      0.086 &      1.397 &       1.14 &      0.120 &      0.098 &      -0.38 &     -0.528 &     -0.389 &      0.148 &      0.009 \\
     10140 &      0.207 &      1.093 &       1.01 &      0.226 &      0.209 &      -0.99 &     -1.429 &     -1.162 &      0.439 &      0.172 \\
     10306 &      0.077 &      0.808 &       1.17 &      0.062 &      0.090 &      -0.38 &     -0.171 &     -0.344 &     -0.209 &     -0.036 \\
     15330 &      0.077 &      1.042 &       1.02 &      0.080 &      0.078 &      -0.20 &     -0.271 &     -0.282 &      0.071 &      0.082 \\
     17147 &      0.137 &      1.113 &       1.05 &      0.152 &      0.144 &      -0.76 &     -0.768 &     -0.673 &      0.008 &     -0.087 \\
     19814 &      0.126 &      1.209 &       1.09 &      0.152 &      0.137 &      -0.70 &     -0.769 &     -0.631 &      0.069 &     -0.069 \\
     21272 &      0.025 &      1.727 &       1.23 &      0.044 &      0.031 &      -0.03 &     -0.078 &     -0.061 &      0.048 &      0.031 \\
     22263 &     -0.022 &      1.307 &       1.00 &     -0.028 &     -0.022 &       0.10 &      0.194 &      0.125 &     -0.094 &     -0.025 \\
     22596 &      0.087 &      1.112 &       1.01 &      0.096 &      0.087 &      -0.32 &     -0.370 &     -0.331 &      0.050 &      0.011 \\
     27913 &      0.027 &      1.163 &       1.01 &      0.031 &      0.027 &      -0.05 &     -0.019 &     -0.044 &     -0.031 &     -0.006 \\
     33495 &      0.121 &      1.135 &       1.14 &      0.138 &      0.138 &      -0.84 &     -0.655 &     -0.637 &     -0.185 &     -0.203 \\
     35377 &      0.107 &      1.055 &       1.02 &      0.113 &      0.109 &      -0.38 &     -0.477 &     -0.453 &      0.097 &      0.073 \\
     36818 &      0.151 &      1.119 &       1.01 &      0.168 &      0.152 &      -0.83 &     -0.901 &     -0.732 &      0.071 &     -0.098 \\
     37853 &      0.155 &      1.105 &       1.02 &      0.171 &      0.158 &      -0.78 &     -0.923 &     -0.773 &      0.143 &     -0.007 \\
     38541 &      0.251 &      1.002 &       1.01 &      0.251 &      0.253 &      -1.76 &     -1.692 &     -1.553 &     -0.068 &     -0.207 \\
     38625 &      0.153 &      1.035 &       1.15 &      0.159 &      0.176 &      -0.93 &     -0.819 &     -0.906 &     -0.111 &     -0.024 \\
     38908 &      0.095 &      1.152 &       1.02 &      0.109 &      0.097 &      -0.36 &     -0.455 &     -0.383 &      0.095 &      0.023 \\
     42438 &      0.073 &      1.115 &       1.01 &      0.082 &      0.074 &      -0.27 &     -0.282 &     -0.260 &      0.012 &     -0.010 \\
     43726 &     -0.006 &      1.466 &       1.06 &     -0.008 &     -0.006 &       0.07 &      0.133 &      0.076 &     -0.063 &     -0.006 \\
     50384 &      0.078 &      1.183 &       1.11 &      0.093 &      0.087 &      -0.38 &     -0.347 &     -0.328 &     -0.033 &     -0.052 \\
     51248 &      0.068 &      1.117 &       1.00 &      0.076 &      0.068 &      -0.23 &     -0.250 &     -0.231 &      0.020 &      0.001 \\
     53070 &      0.184 &      1.094 &       1.13 &      0.202 &      0.208 &      -1.38 &     -1.194 &     -1.159 &     -0.186 &     -0.221 \\
     54772 &      0.197 &      1.075 &       1.17 &      0.211 &      0.230 &      -1.15 &     -1.287 &     -1.344 &      0.137 &      0.194 \\
     56997 &      0.016 &      1.608 &       1.14 &      0.025 &      0.018 &       0.03 &      0.005 &     -0.009 &      0.025 &      0.039 \\
     59750 &      0.128 &      1.135 &       1.13 &      0.146 &      0.145 &      -0.82 &     -0.716 &     -0.682 &     -0.104 &     -0.138 \\
     60632 &      0.227 &      1.018 &       1.17 &      0.231 &      0.265 &      -1.68 &     -1.478 &     -1.666 &     -0.202 &     -0.014 \\
     62207 &      0.095 &      1.152 &       1.03 &      0.110 &      0.098 &      -0.30 &     -0.459 &     -0.392 &      0.159 &      0.092 \\
     63559 &      0.175 &      1.097 &       1.03 &      0.193 &      0.181 &      -0.93 &     -1.111 &     -0.940 &      0.181 &      0.010 \\
     64394 &      0.015 &      1.186 &       1.01 &      0.018 &      0.016 &       0.06 &      0.035 &      0.001 &      0.025 &      0.059 \\
     64426 &      0.121 &      1.123 &       1.08 &      0.136 &      0.131 &      -0.66 &     -0.644 &     -0.589 &     -0.016 &     -0.071 \\
     64924 &      0.002 &      1.629 &       1.10 &      0.004 &      0.002 &      -0.02 &      0.092 &      0.048 &     -0.112 &     -0.068 \\
     69972 &     -0.036 &      1.631 &       2.01 &     -0.059 &     -0.073 &       0.26 &      0.266 &      0.245 &     -0.006 &      0.015 \\
     70681 &      0.218 &      1.077 &       1.00 &      0.235 &      0.218 &      -1.45 &     -1.523 &     -1.243 &      0.073 &     -0.207 \\
     71681 &     -0.010 &      1.652 &       1.47 &     -0.017 &     -0.015 &       0.14 &      0.162 &      0.106 &     -0.022 &      0.034 \\
     71683 &     -0.031 &      1.435 &       1.04 &     -0.044 &     -0.032 &       0.22 &      0.234 &      0.154 &     -0.014 &      0.066 \\
     72998 &      0.136 &      1.143 &       1.14 &      0.155 &      0.155 &      -0.63 &     -0.793 &     -0.751 &      0.163 &      0.121 \\
     73005 &      0.093 &      1.495 &       1.27 &      0.139 &      0.118 &      -0.55 &     -0.665 &     -0.508 &      0.115 &     -0.042 \\
     75181 &      0.110 &      1.075 &       1.02 &      0.119 &      0.113 &      -0.48 &     -0.519 &     -0.476 &      0.039 &     -0.004 \\
     80837 &      0.119 &      1.125 &       1.06 &      0.133 &      0.126 &      -0.64 &     -0.624 &     -0.556 &     -0.016 &     -0.084 \\
     81800 &      0.037 &      1.315 &       1.05 &      0.048 &      0.039 &      -0.01 &     -0.100 &     -0.092 &      0.090 &      0.082 \\
     82636 &      0.105 &      1.046 &       1.03 &      0.110 &      0.108 &      -0.38 &     -0.456 &     -0.447 &      0.076 &      0.067 \\
     84905 &      0.085 &      1.169 &       1.02 &      0.099 &      0.087 &      -0.56 &     -0.389 &     -0.326 &     -0.171 &     -0.234 \\
     88745 &      0.121 &      1.123 &       1.08 &      0.136 &      0.131 &      -0.42 &     -0.644 &     -0.589 &      0.224 &      0.169 \\
     89554 &      0.211 &      1.050 &       1.14 &      0.222 &      0.241 &      -1.44 &     -1.387 &     -1.439 &     -0.053 &     -0.001 \\
     96258 &      0.018 &      0.966 &       1.13 &      0.018 &      0.021 &      -0.13 &      0.038 &     -0.018 &     -0.168 &     -0.112 \\
     96901 &     -0.001 &      1.454 &       1.04 &     -0.001 &     -0.001 &       0.08 &      0.108 &      0.059 &     -0.028 &      0.021 \\
     97063 &     -0.016 &      0.833 &       1.13 &     -0.013 &     -0.018 &       0.02 &      0.149 &      0.114 &     -0.129 &     -0.094 \\
     98020 &      0.208 &      1.091 &       1.00 &      0.227 &      0.208 &      -1.37 &     -1.443 &     -1.160 &      0.073 &     -0.210 \\
     99026 &      0.037 &      0.549 &       1.17 &      0.020 &      0.043 &       0.02 &      0.027 &     -0.112 &     -0.007 &      0.132 \\
     99461 &      0.110 &      1.327 &       1.41 &      0.146 &      0.155 &      -0.58 &     -0.717 &     -0.751 &      0.137 &      0.171 \\
     99889 &      0.050 &      0.642 &       1.19 &      0.032 &      0.059 &      -0.05 &     -0.023 &     -0.185 &     -0.027 &      0.135 \\
    100568 &      0.187 &      1.092 &       1.05 &      0.204 &      0.196 &      -1.22 &     -1.215 &     -1.060 &     -0.005 &     -0.160 \\
    100792 &      0.154 &      1.125 &       1.13 &      0.174 &      0.174 &      -0.99 &     -0.944 &     -0.892 &     -0.046 &     -0.098 \\
    102011 &      0.047 &      0.622 &       1.17 &      0.029 &      0.055 &      -0.03 &     -0.010 &     -0.165 &     -0.020 &      0.135 \\
    102029 &     -0.021 &      0.807 &       1.15 &     -0.017 &     -0.024 &       0.15 &      0.162 &      0.133 &     -0.012 &      0.017 \\
    102485 &     -0.003 &      0.886 &       1.17 &     -0.003 &     -0.004 &      -0.11 &      0.116 &      0.070 &     -0.226 &     -0.180 \\
    102531 &     -0.057 &      0.609 &       1.12 &     -0.035 &     -0.064 &       0.12 &      0.211 &      0.228 &     -0.091 &     -0.108 \\
    103269 &      0.237 &      1.040 &       1.01 &      0.246 &      0.239 &      -1.60 &     -1.638 &     -1.424 &      0.038 &     -0.176 \\
    103498 &      0.171 &      1.099 &       1.08 &      0.188 &      0.185 &      -0.99 &     -1.070 &     -0.971 &      0.080 &     -0.019 \\
    104659 &      0.194 &      1.087 &       1.09 &      0.211 &      0.212 &      -1.42 &     -1.286 &     -1.188 &     -0.134 &     -0.232 \\
    105184 &      0.027 &      1.322 &       1.02 &      0.035 &      0.027 &      -0.14 &     -0.038 &     -0.045 &     -0.102 &     -0.095 \\
    105864 &      0.019 &      0.968 &       1.15 &      0.018 &      0.022 &       0.08 &      0.035 &     -0.023 &      0.045 &      0.103 \\
    109646 &      0.121 &      1.123 &       1.08 &      0.136 &      0.131 &      -0.59 &     -0.644 &     -0.589 &      0.054 &     -0.001 \\
    110778 &      0.071 &      1.116 &       1.01 &      0.079 &      0.071 &      -0.13 &     -0.264 &     -0.246 &      0.134 &      0.116 \\
    110785 &      0.011 &      1.449 &       1.08 &      0.016 &      0.012 &      -0.04 &      0.044 &      0.014 &     -0.084 &     -0.054 \\
    110996 &     -0.032 &      1.638 &       2.27 &     -0.053 &     -0.073 &       0.25 &      0.253 &      0.245 &     -0.003 &      0.005 \\
    113357 &      0.014 &      1.392 &       1.06 &      0.020 &      0.015 &       0.12 &      0.028 &      0.002 &      0.092 &      0.118 \\
    113896 &      0.035 &      1.148 &       1.01 &      0.041 &      0.036 &      -0.10 &     -0.064 &     -0.081 &     -0.036 &     -0.019 \\
    114081 &     -0.011 &      0.853 &       1.15 &     -0.010 &     -0.013 &       0.25 &      0.138 &      0.099 &      0.112 &      0.151 \\
    114096 &      0.037 &      1.315 &       1.05 &      0.048 &      0.039 &       0.09 &     -0.100 &     -0.092 &      0.190 &      0.182 \\
    114210 &      0.048 &      0.629 &       1.18 &      0.030 &      0.056 &      -0.17 &     -0.015 &     -0.173 &     -0.155 &      0.003 \\
    116824 &      0.020 &      0.410 &       1.19 &      0.008 &      0.023 &       0.09 &      0.076 &     -0.029 &      0.014 &      0.119 \\
\hline
\end{tabular}
}  
\end{table*}

%TABLE 6
\begin{table*}
\center
\caption{Statistics for two metallicity calibrations based on new guillotine factors and the ones adopted from \citet{Sandage69}, for two samples: (a) for 75 stars taken from \cite{Karaali03} and (b) for 469 stars taken from \cite{Karatas06}.} 
\begin{tabular}{ccccc}
\hline
 (a)& \multicolumn{2}{c}{Mean deviation $\langle d[Fe/H]\rangle$} & \multicolumn{2}{c}{Standard deviation ($\sigma$)}\\
\hline
$[Fe/H]$(dex) & This paper & \citet{Sandage69} & This paper & \citet{Sandage69} \\
\hline
$-1.76<[Fe/H]\leq-1.0$ & -0.033& -0.123 & 0.116 & 0.139 \\
$-1.0<[Fe/H]\leq-0.5$  & 0.045 & -0.039 & 0.149 & 0.110 \\
$-0.5<[Fe/H]\leq0.0$   & 0.004 &  0.016 & 0.114 & 0.080 \\
$0.0<[Fe/H]\leq0.4$    & 0.009 &  0.054 & 0.081 & 0.084 \\
\hline
    &  &   &  &  \\
\hline
 (b)& \multicolumn{2}{c}{Mean deviation $\langle d[Fe/H]\rangle$} & \multicolumn{2}{c}{Standard deviation ($\sigma$)}\\
\hline
$[Fe/H]$(dex) & This paper & \citet{Sandage69} & This paper & \citet{Sandage69} \\
\hline
$-1.76<[Fe/H]\leq-1.0$ & -0.040& -0.029 & 0.251 & 0.423 \\
$-1.0<[Fe/H]\leq-0.5$  & -0.062 & -0.131 & 0.278 & 0.255 \\
$-0.5<[Fe/H]\leq0.0$   & -0.034 &  0.010 & 0.198 & 0.187 \\
$0.0<[Fe/H]\leq0.4$    & 0.036 &  0.095 & 0.166 & 0.177 \\
\hline
\end{tabular}
\end{table*} 

\section{Summary}

We used the data of 11 authors appearing in the PASTEL catalogue \citep{Soubiran10} and estimated metallicity dependent guillotine factors $f_K$ which are used in an improved metallicity calibration. The metallicities taken from different authors were reduced to the metallicities of \citet{Valenti05}, thus a homogeneous set of metallicities could be obtained. There are differences between the new guillotine factors $f_K$ and the ones $f_S$ adopted from \cite{Sandage69}. 

We derived metallicity calibrations for two sets of guillotine factors using the same procedure and applied them to two different sets of data. The data of the first set were taken from \cite{Karaali03}, whereas the ones of the second set are from \cite{Karatas06}. For the first set, the mean deviations of the residuals for two calibrations are different. The agreement is only for the metallicity interval $-1<[Fe/H]\leq-0.5$ dex, whereas for the metallicity intervals $-1.76<[Fe/H]\leq-1$, $-0.5<[Fe/H]\leq0$ and $0<[Fe/H]\leq+0.4$ dex, the mean deviations corresponding to the metallicity dependent guillotine factors $f_K$ are much smaller than the ones estimated via the guillotine factors adopted from \cite{Sandage69}, $f_S$. Also, the metallicity residuals for the total metallicity interval, $-1.76<[Fe/H]\leq+0.4$, confirms the advantage of the metallicity dependent guillotine factors.

%FIGURE 9
\begin{figure}
\begin{center}
\includegraphics[scale=0.35, angle=0]{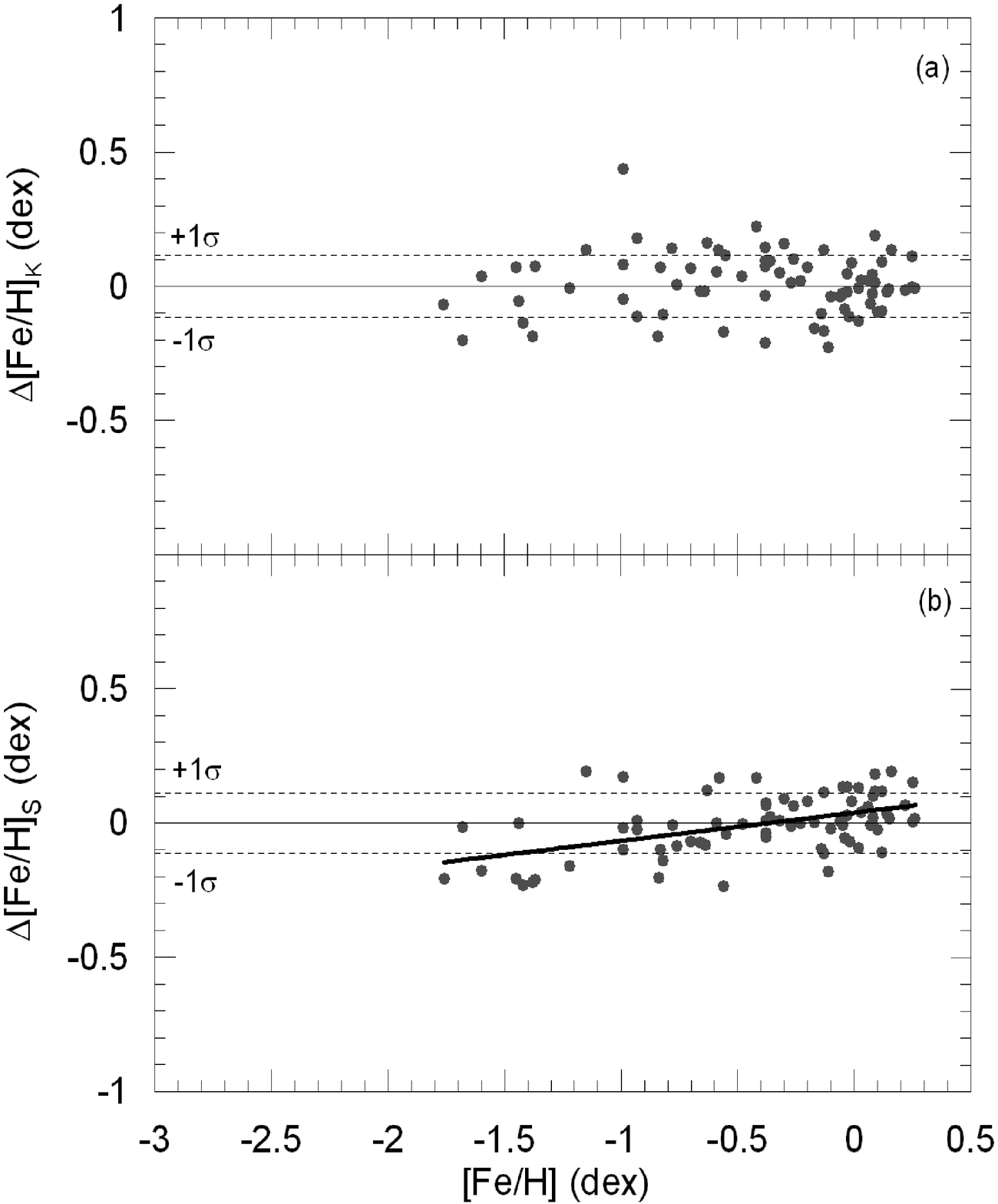}
\caption[] {Metallicity residuals versus metallicity for 75 stars taken from \cite{Karaali03}: for the calibration based on metallicity dependent guillotine factors (a) and the one based on guillotine factors adopted from \cite{Sandage69} (b). The dashed lines denote one standard deviation.}
\end{center}
\end{figure}

%FIGURE 10
\begin{figure}
\begin{center}
\includegraphics[scale=0.37, angle=0]{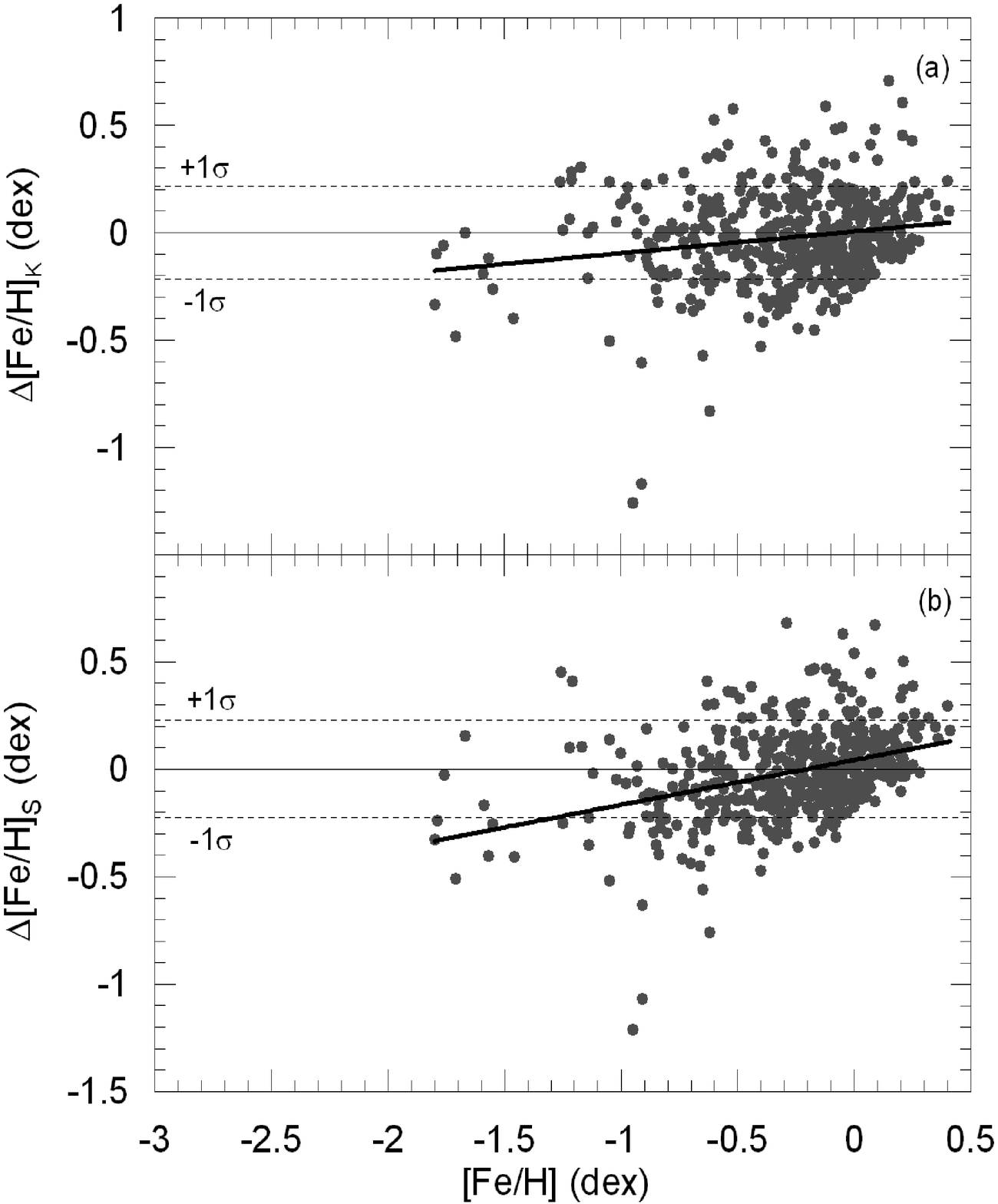}
\caption[] {Metallicity residuals versus metallicity for 469 stars taken from \cite{Karatas06}: for the calibration based on metallicity dependent guillotine factors (a) and the one based on guillotine factors adopted from \cite{Sandage69} (b). The dashed lines denote one standard deviation. The inclination of the calibration line in the upper panel is less than the one in the lower panel, favoring the metallicity calibration based on metallicity dependent guillotine factors.}
\end{center}
\end{figure} 

For the second set, there is an agreement between the mean deviations for two calibrations only for the metallicity interval $-1.76<[Fe/H]\leq-1$ dex. The mean deviation of the residuals estimated via $f_S$, for the metallicity interval $-0.5<[Fe/H]\leq0$ dex, is a bit smaller than the ones estimated via $f_K$, whereas, for two metallicity intervals, i.e. $-1<[Fe/H]\leq-0.5$ and $0<[Fe/H]\leq+0.4$ dex, the mean deviations corresponding to $f_K$ are much smaller than the ones of $f_S$. In Fig. 10, the residuals estimated via $f_K$ and $f_S$ are calibrated to linear equations of the metallicities. However, the inclination of the line for the upper panel (0.10) is less than the one for the lower panel (0.20), indicating that the metallicities estimated by means of the calibration based on metallicity dependent guillotine factors agree better with the original metallicities relative to the other set of estimated metallicities.

We showed that the metallicity dependent guillotine factors provide more accurate metallicities relative to the ones estimated by using the guillotine factors in the literature. This work will be useful for the astronomers who would work with $UBV$ photometry, which has the advantage of being able to be transformed to other systems.

\section{Acknowledgments}
All the authors are greatful to the anonymous referee whose comments and suggestions improved the paper. S. Karaali is grateful to the Beykent University for financial support. This research has made use of the SIMBAD database, operated at CDS, Strasbourg, France and NASA/IPAC Extragalactic Database (NED) which is operated by the Jet Propulsion Laboratory, California Institute of Technology, under contract with the National Aeronautics and Space Administration.

\end{document}